\documentclass{emulateapj}
\usepackage{graphicx} 
  
\def\refjnl#1{{\rm#1}}
\def\aj{\refjnl{AJ}}                   
\def\apj{\refjnl{ApJ}}                 
\def\aap{\refjnl{A\&A}}                
\def\icarus{\refjnl{Icarus}}           
\def\mnras{\refjnl{MNRAS}}             
\def\pasp{\refjnl{PASP}}               
\def\rmxaa{\refjnl{Rev. Mexicana Astron. Astrofis.}} 
\def\sovast{\refjnl{Soviet~Ast.}}      
\def\nat{\refjnl{Nature}}              

\def\be{\begin{equation}} 
\def\ee{\end{equation}}

\def\gsim{\lower.5ex\hbox{\gtsima}} 
\def\lsim{\lower.5ex\hbox{\ltsima}} \def\gtsima{$\; \buildrel > \over \sim \;$} \def\ltsima{$\; \buildrel < \over \sim \;$} \def\prosima{$\; 
\buildrel \propto \over \sim \;$} \def\gsim{\lower.5ex\hbox{\gtsima}} 
\def\lsim{\lower.5ex\hbox{\ltsima}} 
\def\simgt{\lower.5ex\hbox{\gtsima}} 
\def\simlt{\lower.5ex\hbox{\ltsima}} 
\def\simpr{\lower.5ex\hbox{\prosima}}   
  
 \def\gtsima{$\; \buildrel > \over \sim \;$} 
\def\ltsima{$\; \buildrel < \over \sim \;$} 
\def\gsim{\lower.5ex\hbox{\gtsima}} 
\def\lsim{\lower.5ex\hbox{\ltsima}} 
\def\simgt{\lower.5ex\hbox{\gtsima}} 
\def\simlt{\lower.5ex\hbox{\ltsima}} 
\def\simpr{\lower.5ex\hbox{\prosima}}

\def\E3{{\cal E}_{\rm g}^{III}}

\def\Zsun{\rm Z_\odot}

\def\Zsun{\rm Z_\odot}
\def\M*{M_*}
\def\Z*{Z_*}
\def\L*{L_*}

\shorttitle{Most habitable galaxy type}
\shortauthors{Dayal et al.}

\begin{document}

\title{The quest for cradles of life: using the fundamental metallicity relation to hunt for the most habitable type of galaxy}
\author{Pratika Dayal\altaffilmark{1}, Charles Cockell\altaffilmark{2}, Ken Rice\altaffilmark{3} \& Anupam Mazumdar\altaffilmark{4}}
\altaffiltext{1}{Institute for Computational Cosmology, Department of Physics, University of Durham, South Road, Durham DH1 3LE, UK}
\altaffiltext{2}{UK Centre for Astrobiology, School of Physics and Astronomy, University of Edinburgh, EH9 3HJ, UK}
\altaffiltext{3}{Institute for Astronomy, University of Edinburgh, Royal Observatory, Edinburgh, EH9 3HJ, UK}
\altaffiltext{4}{Consortium for Fundamental Physics, Lancaster University, Lancaster, LA1 4YB, UK }

\begin{abstract}
The field of astrobiology has made huge strides in understanding the habitable zones around stars (Stellar Habitable Zones) where life can begin, sustain its existence and evolve into complex forms. A few studies have extended this idea by modelling galactic-scale habitable zones (Galactic Habitable Zones) for our Milky Way and specific elliptical galaxies. However, estimating the habitability for galaxies spanning a wide range of physical properties has so far remained an outstanding issue. Here, we present a ``cosmobiological" framework that allows us to sift through the entire galaxy population in the local Universe and answer the question ``{\it Which type of galaxy is most likely to host complex life in the cosmos"}? Interestingly, the three key astrophysical criteria governing habitability (total mass in stars, total metal mass and ongoing star formation rate) are found to be intricately linked through the ``fundamental metallicity relation" as shown by SDSS (Sloan Digital Sky Survey) observations of more than a hundred thousand galaxies in the local Universe. Using this relation we show that metal-rich, shapeless giant elliptical galaxies at least twice as massive as the Milky Way (with a tenth of its star formation rate) can potentially host ten thousand times as many habitable (earth-like) planets, making them the most probable ``cradles of life" in the Universe. \\
\end{abstract}


\section{Introduction}

A key question in astrobiology is ``what is the distribution of habitable conditions and life in the Universe"?

Observational astronomers have concentrated their effort on peering into the skies hunting for earth-like ``exoplanets" (see e.g. Seager 2010) and amassed a database numbering in the thousands\footnote{database available at http://exoplanets.org}. In unison, theorists have made huge strides in defining the requirements for habitable zones (where planets can produce biospheres) to exist around stars of varying ages, masses and metallicities\footnote{In astrophysics, metals denote any element with a higher atomic number than Hydrogen}, the stellar habitable zones (SHZ; e.g. Huang 1959; Hart 1979; Kasting et al. 1993; Rushby et al. 2013; Kopparapu et al. 2013; G\"uedel et al. 2014). A few theoretical works have extended SHZ calculations to estimate habitable zones at galactic scales, the Galactic Habitable Zone (GHZ; Gonzalez et al. 2001) for specific galaxies like the Milky Way (MW) and nearby ellipticals (Suthar \& McKay 2012; Carigi \& Meneses-Goytia 2013). Approaches to modelling the GHZ of the MW range from using metallicity and star formation distributions (Gonzalez et al. 2001) to sophisticated descriptions of the distribution of terrestrial planets (Lineweaver et al. 2004) to modelling individual star systems (Gowanlock et al. 2011) to tracing the co-evolution of the GHZ and the assembly of the MW (and Andromeda) using cosmological simulations (Forgan et al. 2015, submitted to International Journal of Astrobiology). However, estimating the habitability of a whole galaxy population spanning a wide range of physical properties has so far remained an outstanding issue.

In this work, we present the first (and simplest) ``cosmobiological" (Bernal 1952; Dick 1996) formalism that can sift through the entire population of galaxies in the local Universe to answer the question ``{\it Which type of galaxy is most habitable in terms of complex life in the Universe?}"

We draw on the progress made by SHZ and GHZ studies to isolate the primary physical ingredients required to model habitability: planets form in stellar vicinities with a metallicity-dependent probability since planet-formation requires elements heavier than iron. Further, explosions of Type II supernovae (SN henceforth) can expose surfaces of nearby earth-like planets to high doses of cosmic radiation, potentially causing partial or mass extinctions or delaying the evolution of complex life. Although SN may not sterilise a planet, planets in proximity to regular SN may be less habitable with respect to complex life (Clark et al. 1977). We use this understanding to propose three key criteria determining the existence of habitable oases at the galactic scale: the total number of stars, their metallicity and the probability of planets existing, unmolested by poisonous SN radiation. Naturally, galaxies with the largest number of potentially habitable planets would be most likely to be ``habitable". 

\newpage 
\section{The Model}

We start by linking the habitability criteria mentioned above to observable astrophysical quantities: firstly, the total number of stars scales with the total mass in stars (stellar mass) in a galaxy. Secondly, we make the reasonable assumption that the metallicity of stars closely follows that of the gas from which they formed, with the gas-phase metallicity being a direct observable for galaxies in the local Universe. Thirdly, we make the simplifying assumption that stars are homogeneously distributed throughout the galaxy so that the probability of planets hosting complex life is inversely proportional to the fractional volume irradiated by recent SN: a SN exploding in a large galaxy will have far less of an impact on far-flung planets as compared to a small galaxy where a few SN could expose all of the planets to high doses of cosmic radiation, rendering them less hospitable for complex life. The fractional volume irradiated by SN depends on the volume affected by a single SN multiplied by the SN rate, divided by the total volume of the galaxy. While some theoretical estimates suggest only SN going off closer than 10 parsecs (pc)\footnote{A parsec is a cosmological distance measure equalling $3 \times 10^{18}$ cm.} to a planet can harm complex life or cause mass extinctions (Ellis \& Schramm 1995), the effects of high radiation doses and climatic disturbances on complex life are only poorly understood at best (Scalo \& Wheeler 2002). We encode our uncertainty by defining that only SN going off closer than $Q$ pc to a planet can potentially harm/inhibit the growth of complex life which yields a total affected volume of $\frac{4}{3} \pi (Q \, {\rm pc})^3$. Type II SN explode when the core mass exceeds the Chandrasekhar mass limit of 1.44 solar masses producing about $10^{53}$ erg/s of energy, 90\% of which is radiated away as cosmic radiation (Shklovskii 1960; Woosley \& Janka 2005) that could be potentially harmful for complex life. Given that all SN radiate roughly the same amount of energy in harmful radiation, $Q$ can be assumed to be the same for all SN which results in it being factored out in our final equations; the exact value of this unknown parameter then never enters into our calculations which focus on defining the habitability of galaxies {\it relative to the MW}. We note that this is an upper limit to the total volume irridiated since SN generally go off in high-density star-formation regions irradiating the {\it same} volume. Further, the SN rate is linked to the ongoing star formation rate through the initial mass function (IMF) that describes the distribution of mass in a freshly formed stellar population. For this work, we assume the commonly-used Salpeter IMF (Salpeter 1955) for stars in the range of $1-100$ solar masses ($M_\odot$), with stars more massive than $8M_\odot$ exploding instantaneously as SN; the total volume of a galaxy roughly scales with the total stellar mass (e.g. Moster et al. 2010) and we ignore the scatter in this relation for simplicity. 

This yields the three key astrophysical criteria that determine the habitability of a galaxy : (a) the total number of stars ($N_{tot}$); (b) the probability of these stars hosting terrestrial planets or gas-giants ($P_t$ and $P_g$ respectively); and (c) the fraction of these planets that are unaffected by SN radiation and therefore have a higher probability of supporting complex life ($P_{cl}$). Each of these terms is derived as now explained in what follows.

The total number of stars ($N_{tot}$) present in a galaxy should be proportional to the total stellar mass ($M_*$) built up over its entire assembly history such that $N_{tot}\propto M_*$.

\begin{figure}
 \center{\includegraphics[scale=0.47]{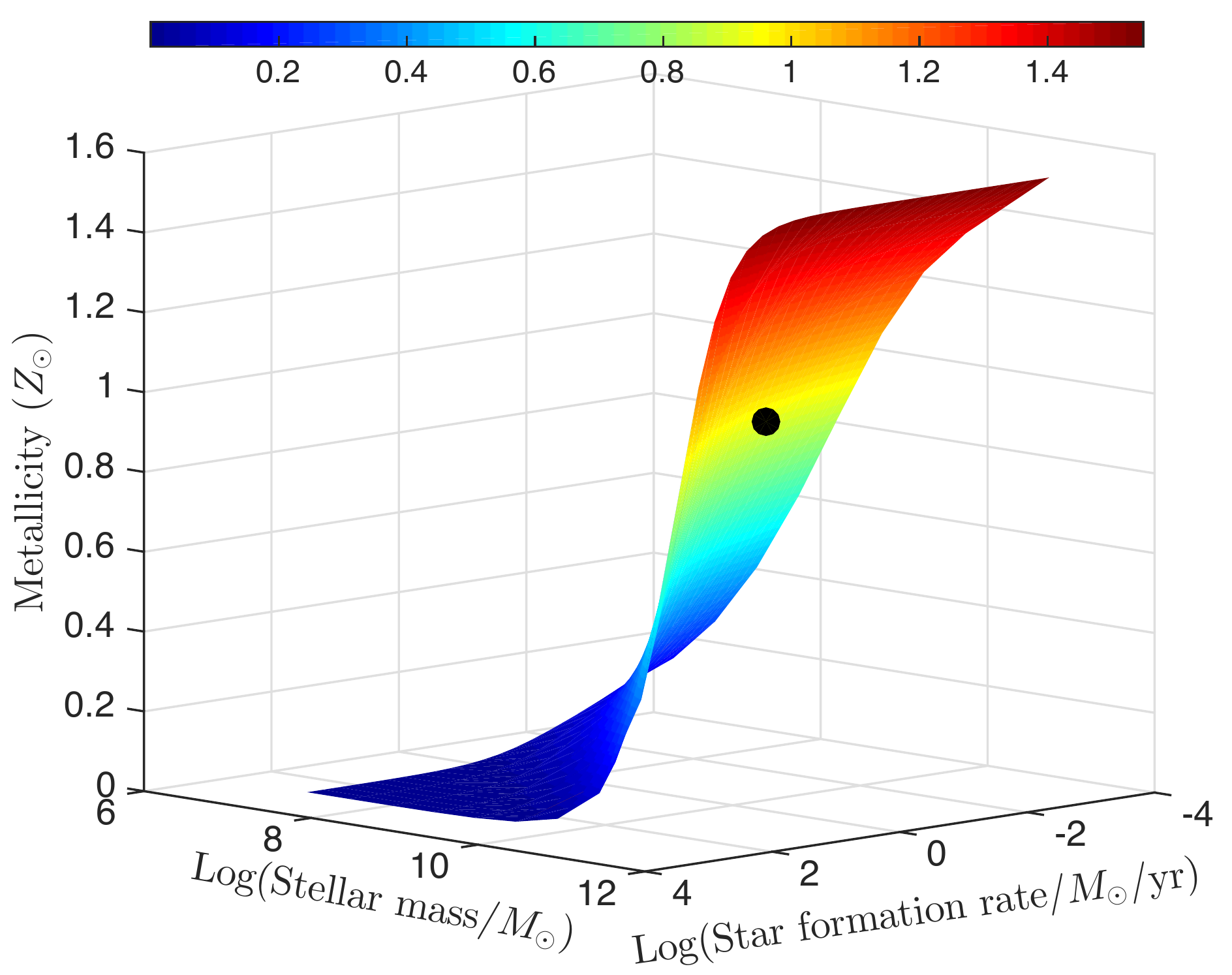}}
   \caption{The fundamental metallicity relation (FMR) linking the key physical properties of total stellar mass, star formation rate and gas-phase metallicity (Mannucci et al. 2010; Lara-lopez et al. 2010). The regions are colour-coded according to the values of the absolute metallicity shown by the colour bar. Sloan Digital Sky Survey (SDSS) observations show that more than a hundred thousand galaxies in the local Universe lie on a very thin plane (with minimal dispersion) in this three dimensional space, showing an intricately linked physical origin. Our Milky Way (black dot) is a ``typically average" galaxy, as shown by its position on the plane.  }
   \label{fig_fmr}
\end{figure}

Next, we link the probability of the galaxy having terrestrial ($P_t$) or gas-giant planets ($P_g$), depending on its metallicity where we implicitly assume that gas and stellar-phase metallicities are correlated. Buchhave et al. (2012) have used Kepler data for 226 exoplanets and spectroscopy of their host stars to show that small terrestrial planets (with radius less than 4 times the earth radius) might be equally probable around stars with metallicities ranging between $0.25-2.5$ solar metallicity ($\Zsun$; see also Petigura et al. 2013). Given the small sample, we assume that the probability of finding terrestrial planets can be related to the gas-phase metallicity ($Z_g$) as $P_t \propto Z_g^\alpha$, where $\alpha$ is a free parameter. We then study $P_t$ in two limiting scenarios: $\alpha=0$ ($\alpha=1$) such that there is no (a strong) probability dependence of terrestrial planets on the gas-phase metallicity; while the first agrees better with observations (Buchhave et al. 2012; Petigura et al. 2013), the latter gives an indication of the variation expected if the probability were to depend strongly on metallicity. To obtain $P_g$, we use the results of Wang \& Fischer (2015) who have obtained high-resolution spectroscopy of about 1040 stars in the Keck, Lick and Anglo-Australian Telescope planet search surveys to show that the probability of gas-giants scales with the metallicity as $P_g \propto 10^{2 Z_g}$.

\begin{figure*}
 \center{\includegraphics[scale=0.52]{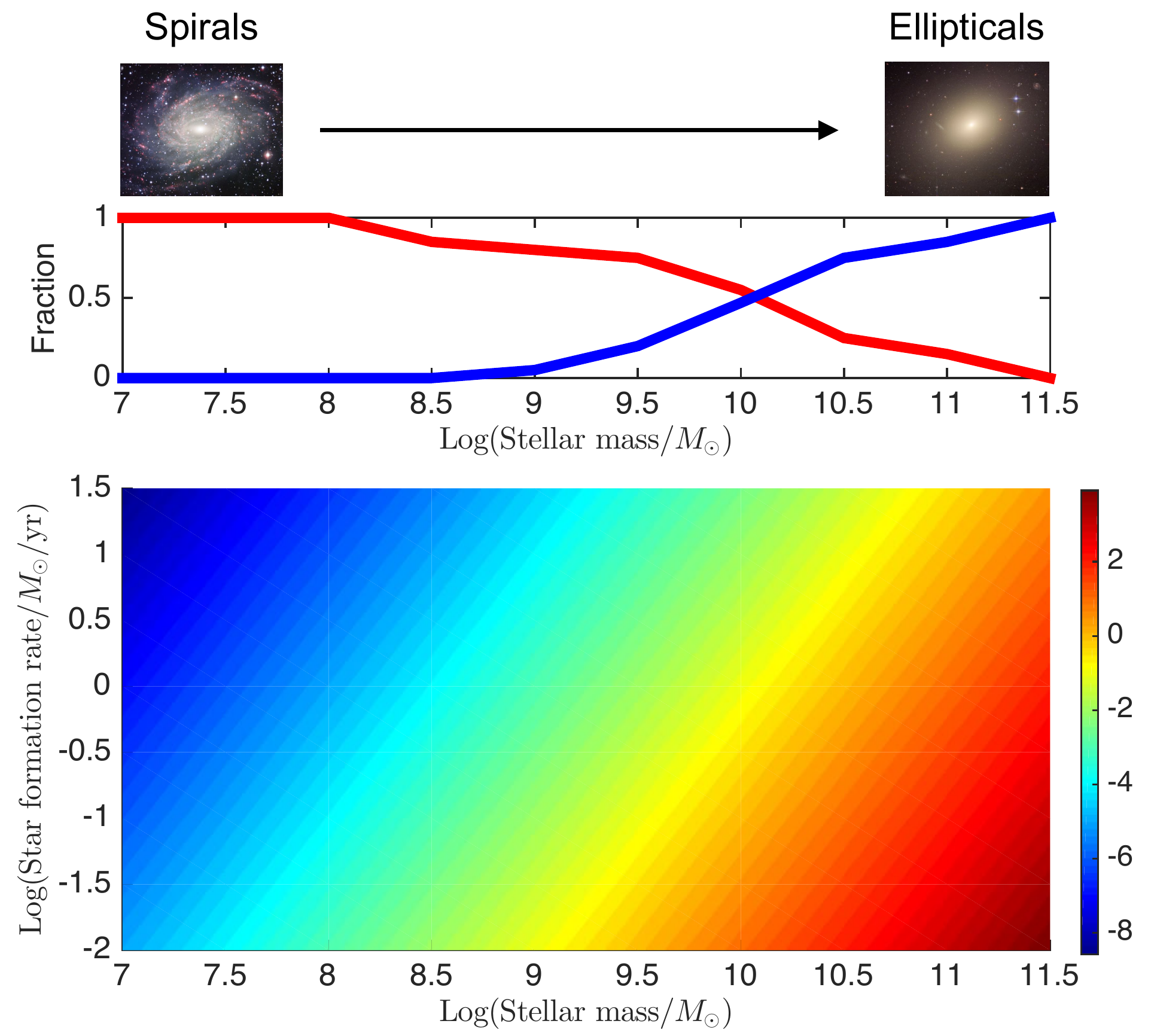}}
   \caption{{\it Upper panel:} The fraction of total galaxies that can be identified as well-defined spirals (red line) and blob-like ellipticals (blue line) as a function of the stellar mass, as observed by the Galaxy And Mass Assembly (GAMA) survey (Kelvin et al. 2014). While low mass galaxies are most often spiral systems their fraction drops with mass to roughly half at $10^{10} M_\odot$ with the most massive galaxies ($\gsim 10^{11} M_\odot$) almost all being ellipticals. {\it Lower panel:} The number of earth-like habitable planets in galaxies occupying different regions of the star formation rate-stellar mass plane, normalised to the MW assuming no-metallicity dependence ($\alpha=0$ in Eqn. 6). The values of the ratio (in log) are shown by the colour bar. Our results clearly show that the number of habitable earth-like planets (compared to the MW) increases on the diagonal that traces increasing stellar mass and decreasing star formation rate. Low-mass spirals ($\lsim 10^{9} M_\odot$) of any star formation rate provide inhospitable environments for earth-like planets to form and evolve. It is predominantly giant ellipticals (masses larger than 2.5 times the MW) with low star formation rates (less than a tenth of the MW) that provide the best environment for habitable planets to form.}
   \label{fig_habt}
\end{figure*}

Finally, we need to calculate the fraction of planets that receive high doses of cosmic radiation due to nearby SN. SN explode after roughly 28.6 million years (Myrs; Padovani \& Matteucci 1993) and so the volume irradiated by a SN due to the ongoing star formation ($\psi$) is
\begin{equation}
V_{irr} = \frac{\psi}{53.17}  \frac{4}{3} \pi (Q\, {\rm pc})^3 (28.6 \,{\rm Myrs})
\end{equation}
where $[53.17M_\odot]^{-1}$ is the SN rate for the chosen Salpeter IMF and SN can sufficiently enhance ionizing radiation so as to affect planets within a radial value of $Q$ pc as explained above. Further, the total volume occupied by the stars is taken to be $V_{tot} = M_*/\rho$ where we assume a homogeneous stellar-mass distribution. The interstellar media of galaxies is assumed to have a density $\rho = 200 \rho_{crit}$ where $\rho_{crit}$ is the critical density of the Universe. Then, the factional volume irradiated by recent SN is
\begin{equation}
f_{irr} = \frac{V_{irr}}{V_{tot}} = \frac{(\psi/53.17) \times (4/3) \pi (Q \, {\rm pc})^3 (28.6 \,{\rm Myrs})}{M_*/\rho}
\end{equation}
The probability of planets suitable for hosting complex life ($P_{cl}$) is then inversely proportional to the volume irradiated by recent SN as explained above.

Therefore, the number of habitable terrestrial planets will be some fraction of those that have avoided SN irradiation, and can be expressed as 
\begin{equation}
N_t \propto N_{tot} P_t P_{cl} \propto \frac{N_{tot} Z_g^\alpha}{f_{irr}}  
\end{equation}
or,
\begin{equation}
N_t \propto \frac{0.739 M_* Z_g^\alpha M_*/\rho}{(\psi/53.17) \times (4/3) \pi (Q\, {\rm pc})^3 (28.6 \, {\rm Myrs})}.
\end{equation}
Removing the various constants, 
\begin{equation}
N_t \propto \frac{M_*^2 Z_g^\alpha}{\psi}.
\end{equation}

Using the same approach, the number of habitable gas-giants can be expressed as
\begin{equation}
N_g \propto \frac{M_*^2 10^{Z_g}}{\psi}.
\end{equation}
Of course, we don't necessarily mean that the gas-giants themselves could be habitable, but that they may host moons/satellites that could have conditions suitable for life (Kipping et al. 2009).

\section{Linking habitability with astrophysical observations}

In an exciting development, observations with the Sloan Digital Sky Survey (SDSS) show that the three key physical properties (total stellar mass, ongoing star formation rate and gas-phase metallicity) are intricately linked with minimal scatter for more than a 140,000 galaxies in the local Universe through an intrinsic ``Fundamental Metallicity Relation" (FMR; Mannucci et al. 2010; Lara-lopez et al. 2010) as shown in Fig. \ref{fig_fmr}. The same figure shows that the Milky Way is a typical galaxy, lying near the knee of the stellar mass-star formation rate-metallicity plane with values of $M_* \simeq 6 \times 10^{10} M_\odot$, $\psi = 3 M_\odot {\rm yr^{-1}}$ and $Z_g =Z_\odot$. It is now understood that this relation emerges as a consequence of the ejection of metal-rich gas and the accretion of metal poor gas over the entire lifetime of any galaxy (Dave et al. 2012; Dayal et al. 2013; Lilly et al. 2013): the position of a galaxy on this relation is therefore a ``fingerprint" of its entire life-history. Given that the metallicity is a function of the stellar mass and star formation rate, the total number of habitable terrestrial planets and gas-giants can also be expressed solely in terms of these two quantities. We then obtain the expected number of habitable planets for about 140,000 SDSS galaxies by scaling their stellar mass and star formation rate compared to the MW as a reference.

Our approach has a two-fold strength: firstly, as a consequence of using the FMR our model implicitly includes the entire formation history of all the galaxies in the local Universe, a critical necessity for modelling GHZ (Gonzalez et al. 2001; Lineweaver et al. 2004; Gowanlock et al. 2011). Secondly, by using the MW as a ``reference" for habitability, we are able to remove all the constants of proportionality to obtain Eqns. 6 and 7.

\begin{figure} 
   \center{\includegraphics[scale=0.47]{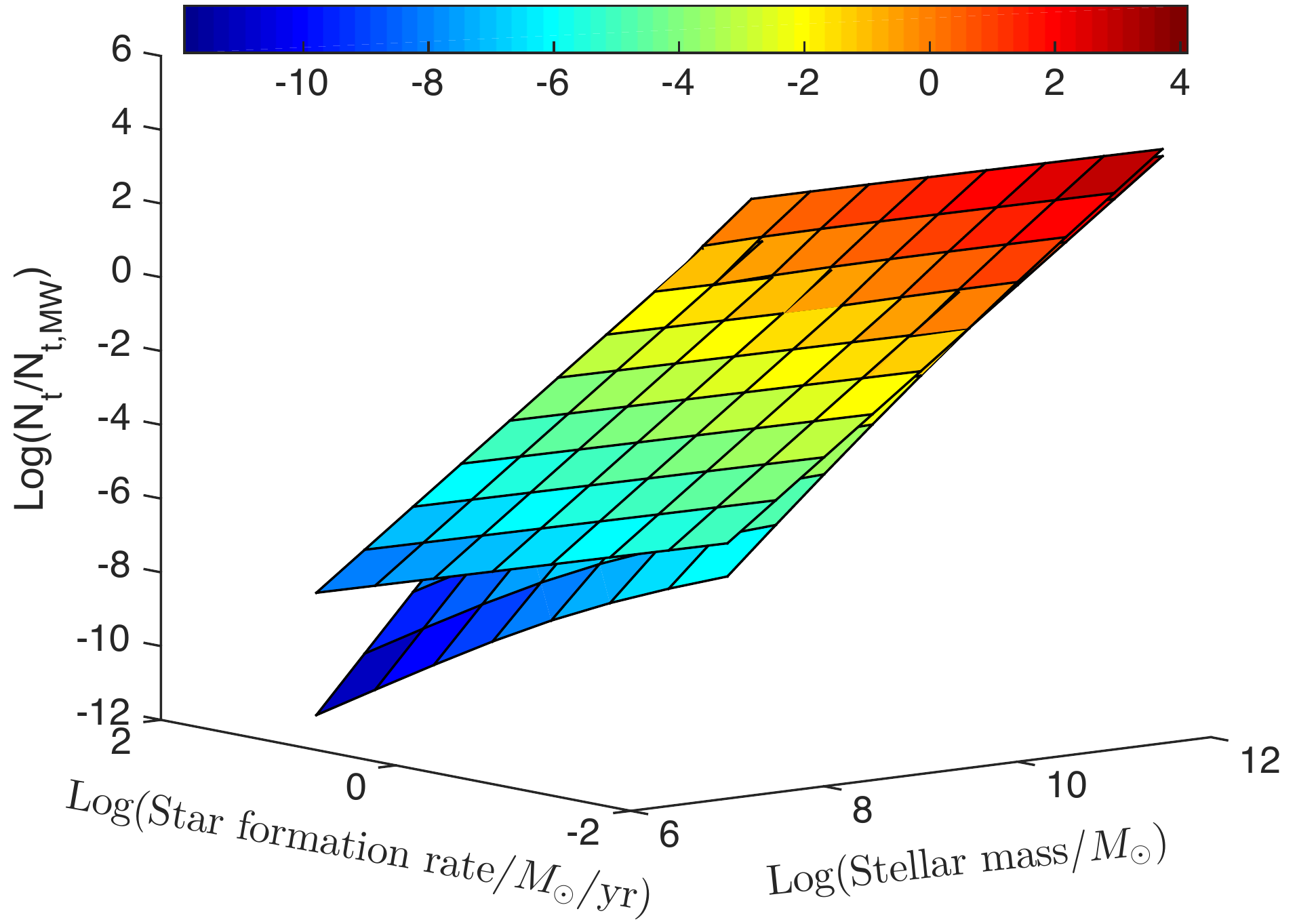}}
  \caption{The number of earth-like habitable planets in galaxies occupying different regions of the star formation rate-stellar mass plane, normalised to the MW. The regions are colour-coded according to the values of the ratio (in log) shown by the colour bar. The two surfaces in this plot are for two different assumptions linking the planet-formation probability with the metallicity: the upper surface shows results for no-metallicity dependence (i.e. $\alpha=0$) in Eqn. 6 while the lower surface shows results for a very strong dependence on metallicity ($\alpha=1$) in the same equation. As seen, the number of planets decreases by about 10,000 assuming a very-high metallicity dependence in low mass galaxies. With their high metal enrichment, larger galaxies are rather insensitive to metallicity.}
  \label{fig_habt2}
\end{figure}

  \begin{figure*}
 \center{\includegraphics[scale=0.65]{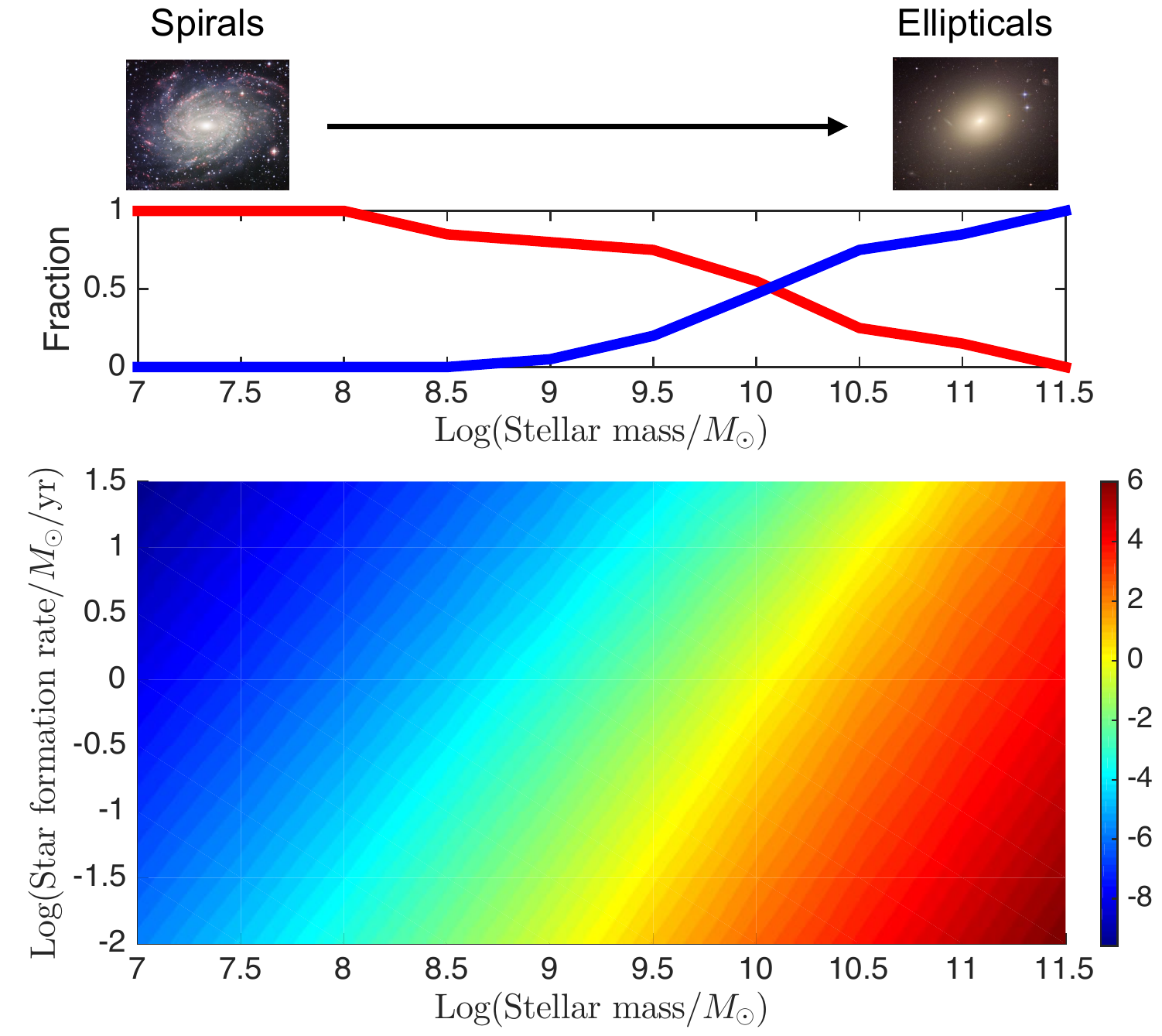}}
   \caption{{\it Upper panel:} The fraction of total galaxies that can be identified as well-defined spirals (red line) and blob-like ellipticals (blue line) as a function of the stellar mass, as observed by the GAMA survey and explained in Fig. \ref{fig_habt}. {\it Lower panel:} The number of Neptune-like (gas giant) planets in galaxies occupying different regions of the star formation rate-stellar mass plane, normalised to the MW. The regions are color-coded according to the values of the ratio (in log) shown by the colour bar. As for earth-like planets, low-mass spirals ($\lsim 10^{9} M_\odot$) with any star formation rate provide inhospitable environments for gas-giants to form and evolve. It is predominantly giant ellipticals (masses larger than 2.5 times the MW) with low star formation rates (less than a tenth of the MW) that provide the best environment for gas-giants to form.}
   \label{fig_habg}
\end{figure*}

\section{Results and discussion}
Our results show that compared to the MW, the number of habitable terrestrial planets increases with {\it increasing stellar mass and decreasing star formation rate}, assuming the simplest case of no-metallicity dependence (i.e. $\alpha=0$ in Eqn. 1), as shown in Fig. \ref{fig_habt}. Physically, this is sensible because the number of stars (and hence planets) increases with the stellar mass, with a low star formation ensuring most of these systems are unaffected by harmful SN radiation. On the other hand, low mass galaxies will host a comparatively small number of planets, most of which will be rendered less hospitable by SN. To quantify, low-mass galaxies ($\lsim 10^{9}$ solar masses) with {\it any} star formation rate provide less hospitable environments compared to the MW, typically hosting {\it a hundred to ten-million times less} earth-like planets capable of supporting complex life. Its is predominantly massive galaxies (masses larger than twice the MW) with less than a tenth of the star formation rates that provide the best environment for habitable planets hosting complex life. 

We then use results from the Galaxy And Mass Assembly (GAMA) survey (Kelvin et al. 2014) which is the largest study linking the stellar mass to the morphology (shape) of local Universe galaxies. The results of this survey (upper panel of Fig. \ref{fig_habt}) show that low mass ($\lsim 10^9$ solar masses) galaxies are most often spiral systems with a fraction that drops to roughly half at a stellar mass of $10^{10}$ solar masses. However, the most massive galaxies ($\gsim 10^{11}M_\odot$) are predominantly all ellipticals, with a fraction that drops steeply with decreasing mass. Therefore our results show that {\it giant ellipticals} with mass $\gsim 10^{10.5} M_\odot$ and star formation rates $\lsim 0.3 M_\odot \, yr^{-1}$ offer the most hospitable environments for planets to form and can host {\it 100-10000 times more earth-like planets than the MW}.

The lowest mass galaxies ($M_* \lsim 10^8$ solar masses) with the largest star formation rates ($\gsim 1$ solar mass per year) are the most metal poor in the local Universe as a result of SN pushing out most of their metals. A strong metallicity dependence ($\alpha = 1$ in Eqn. 1) has the largest impact on these tiny metal-depleted systems: while the the number of habitable terrestrial planets decreases by about 10,000 for these lowest mass galaxies as compared to no-metallicity dependence, massive metal-rich galaxies are effectively insensitive to the metallicity term, as shown in Fig. \ref{fig_habt2}.

Similar to the behaviour shown for earth-like planets, low-mass spirals ($\lsim 10^{9} M_\odot$) of any star formation rate provide less hospitable environments for gas-giants to form and evolve as shown in Fig. \ref{fig_habg}. It is predominantly giant ellipticals (masses larger than twice the MW) with a tenth of its star formation rates that provide the best environment for gas-giants that can host complex life. Indeed, the most-massive lowest-star forming giant elliptical galaxies can host as many as 1,000,000 gas-giants compared to the MW.

To conclude, our quest was to answer the question, ``Which type of galaxy is most habitable"? Sifting through a sample of more than 140,000 local Universe galaxies, we find that the most habitable galaxies are {\it giant ellipticals with masses larger than twice that, and star formation rates less than a tenth of the MW. These shapeless cosmic giants can host {\it ten-thousand} times as many earth-like planets and {\it a million times as many} gas-giants capable of supporting complex life as compared to the MW, making them the sites of the largest number of potentially habitable or life-bearing planets in the Universe.}

The closest giant elliptical to the milky Way is Maffei1 (Maffei 1968) which was discovered in 1968 by the Italian astronomer Paolo Maffei. The latest measurements (Fingerhut et al. 2003) suggest this galaxy lies at a distance of 2.92 Mpc (where a Mpc = $3.086 \times 10^{24}$cm) from our galaxy: if we were listening to our cosmic neighbours, any light signal transmitted from Maffei1 would take about 9.2 Million years to reach us.

We end by noting that our endeavour has been to present the simplest cosmobiological formalism to hunt for the most habitable type of galaxy. Whilst our approach accounts for the entire formation history of all the galaxies in the local Universe, a critical necessity for modelling GHZ, we have naturally made a number of simplifying astrophysical assumptions and approximations: we assume stars (and hence SN) are homogeneously distributed in galaxies, the gas and stellar metallicities trace each-other and that the volume scales with the total stellar mass. Finally, the effects of SN cosmic radiation in causing partial or mass extinctions or delaying the evolution of complex life are only poorly understood at best. We recognise all these assumptions have ample scope for further improvement which we aim to tackle in future studies.

\section*{Acknowledgments} 
PD thanks the Adisson Wheeler fellowship and the Institute for Advanced Study at Durham University for allowing her to undertake this project. PD thanks D. Forgan and N. libeskind for illuminating discussions.



\label{lastpage} 
\end{document}